\documentclass[aps,prl,reprint,nofootinbib,twocolumn,superscriptaddress,showpacs,showkeys,longbibliography]{revtex4-1}
\usepackage{eurosym}
\usepackage{amsmath,amssymb,amstext}
\usepackage{ulem} 
\usepackage[usenames,dvipsnames]{color}
\usepackage{graphicx}
\usepackage{braket}
\usepackage{natbib}
\usepackage{comment}
\usepackage{dcolumn}
\usepackage[english]{babel}
\usepackage{wasysym}
\usepackage[colorlinks,bookmarks=false,citecolor=blue,linkcolor=red,urlcolor=blue]{hyperref}

\begin{document}
\title{Beliaev Damping in Spin-$\frac{1}{2}$ Interacting Bosons with Spin-Orbit Coupling}
\author{Rukuan Wu,}
\affiliation{Department of Physics, Zhejiang Normal University, Jinhua, 321004, China}
\author{Zhaoxin Liang}
\email{The corresponding author: zhxliang@gmail.com}
\affiliation{Department of Physics, Zhejiang Normal University, Jinhua, 321004, China}

\date{\today}

\begin{abstract}
Beliaev damping provides one of the most important mechanisms for dissipation of quasiparticles through beyond-mean-field effects at zero temperature. Here we present the first analytical result of Beliaev damping in low-energy excitations of spin-$\frac{1}{2}$ interacting bosons with equal Rashba and Dresslhaus spin-orbit couplings. We identify novel features of Beliaev decay rate due to spin-orbit coupling, in particular, it shows explicit dependence on the spin-density interaction and diverges at the interaction-modified phase boundary between the zero-momentum and plane-wave phases. This represents a manifestation of the effect of spin-orbit coupling in the beyond-mean-field regime, which by breaking Galilean invariance couples excitations in the density- and spin-channels. By describing the Beliaev damping in terms of the observable dynamic structure factors, our results allow direct experimental access within current facilities. 
\end{abstract}

\maketitle

The dissipation of quasiparticles through their mutual interactions lies at the foundational aspect of the quantum many-body physics \cite{WenBook}, which understanding provides crucial insights into the beyond-mean-field effects of the system. A paradigmatic example is the Beliaev damping \cite{Beliaev1,Beliaev2} of Bose superfluid \cite{PineBook}, where a quasiparticle disintegrates into two quasiparticles even at zero-temperature by colliding with the condensate. After the first realization of Bose-Einstein condensate (BEC) \cite{BEC1995-1,BEC1995-2}, experiments probing the linear Bogoliubov's mode \cite{Kurn1999} and Beliaev damping \cite{Beliaev2001E,Beliaev2002E,Ozeri2005} were immediately carried out. Recently, Beliaev damping of quasiparticles in various exotic superfluids has attracted significant interests, e.g., in the mixture of BECs with normal Fermi gas \cite{Yip2001,Santamore2004,Santamore2005,Liu2003}, the Fermi superfluids \cite{Zheng2014,Shen2015,Pixley2015}, the dipolar BECs \cite{Natu2013A,Natu2013B,Natu2016} and the non-equilibrium polariton BECs \cite{Regemortel2017}.

This work is motivated by current experimental progress highlighting realizations of spin-orbit coupling (SOC) with ultracold quantum gases \cite{Lin2009,Lin2011,Wangpj2012,Cheuk2012,Zhang2012,Ji2014,Li2016,Li2017}, which opens new routes toward exotic quantum many-body systems in gauge fields \cite{Dalibard2011,Galitski2013,Zhai2012,Zheng2013,Goldman2014,Zhai2015,Zhang2016}. The SOC, where the motion of particles are coupled to their spin, breaks the Galilean invariance, giving rise to a double-minimum single-particle energy spectrum. Thus a SOC BEC has the crucial novelty already at the mean-field level compared to the SOC-free counterpart, as have been intensively studied \cite{Lin2009,Lin2011,Wangpj2012,Cheuk2012,Li2016,Li2017,Dalibard2011,Galitski2013,Zhai2012,Zhai2015,Zhang2016}, in particular, (i) In the ground state, an exotic stripe phase \cite{Wang2010,Li2012,Li2013,Li2017} spontaneously breaking translational symmetry can emerge; (ii) For non-interacting quasiparticles, a softening of phonon or roton modes occurs \cite{Chen2015}, and more importantly, the critical superfluid velocity cannot be well defined \cite{Zhu2012,Zhu2015,Zhai2015} without {\it a priori} choice of reference frame. Beyond the mean field, however, the consequence of SOC coupling the spin and motional degrees of freedom on the dissipation of quasiparticles, such as Beliaev damping, remains elusive and challenging.

In this Letter, we present the first analytical result on the Beliaev decay of phonons in a SOC BEC [see Eq.~(\ref{BF})] allowing insights into the interplay between SOC and {beyond-mean-field} effects. Considering the condensate in zero-momentum phase \cite{Lin2011}, we find that the damping of phonons, while maintaining the familiar $q^5$ scaling with momenta, exhibits two novel features in contrast to the SOC-free counterpart. First, the damping rate becomes explicitly dependent on the interaction constant, to be precise, the strength of the spin-density interaction. Remarkably, the damping rate diverges at the critical point which exactly corresponds to the interaction-modified phase boundary between the plane-wave and zero-momentum phases. Second, the damping of phonons becomes strongly anisotropic. The former is a result of SOC coupling the density- and spin-density excitations due to absence of Galilean invariance, while the latter is a manifestation of the SOC-induced anisotropic effective mass. Our work will shed light on the understanding of dissipation of elementary excitations in non-Abelian gauge field at zero temperature.

\begin{figure}[!ht]
\centering
\includegraphics[width=1.01\columnwidth]{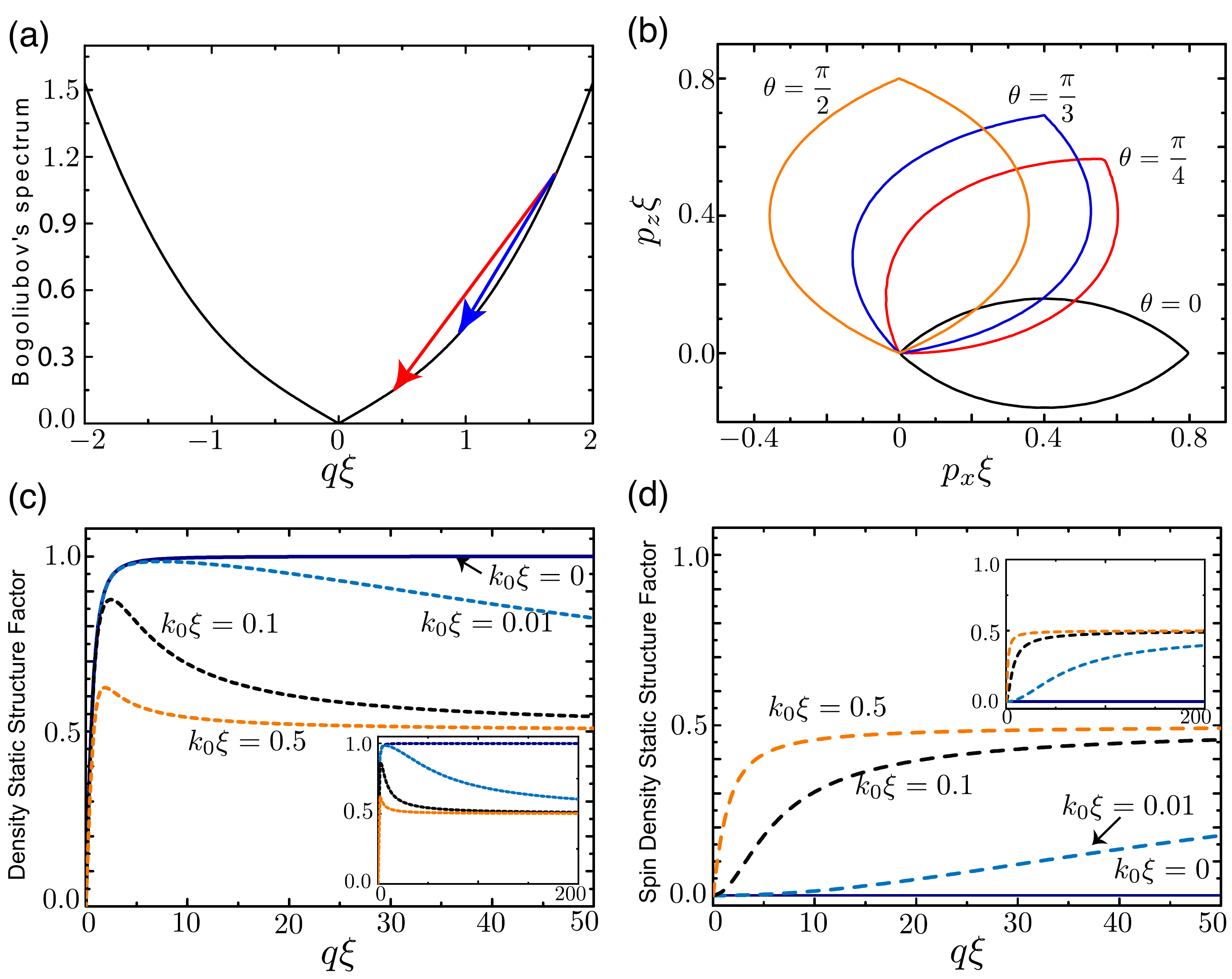}
\caption{(color online). (a) Dispersion of density mode in the SOC BEC [see Eq.~(\ref{Ham})] in the zero momentum phase. Arrows schematically show the Beliaev decay of a Bogoliubov mode with momentum ${\bf q}$ into two modes with momenta ${\bf p}$ and ${\bf q-p}$, respectively. (b) The momentum $\textbf{p}$ manifold allowed by the energy and momentum conservations in $p_x-p_z$ plane, considering various direction of the initial momentum ${\bf q}$. Specifically, we fix the modulus of ${\bf q}$ as $q\xi=0.8$ while varying its angle $\theta$ with respect to the SOC direction along $x$ axis. For SOC strength, we take $k_0\xi=0.5$. (c) Density and (d) spin-density static structure factor as a function of $q$ for various $k_0$. Here, $\theta=0$ is taken for illustration. Insets plot the asymptotic behavior of corresponding static structure factors at large momenta. In all plots, the momentum is measured in units of the inverse coherence length ($\xi^{-1}=\sqrt{\hbar/m\Omega}$). For other parameters, we take $G_1/\hbar\Omega = 0.1$ and $G_2/\hbar\Omega = 0.025$.  }\label{figure1}
\end{figure}

{\it Beliaev Damping---} For a quasiparticle carrying momentum ${\bf q}$, its decay rate in the Beliaev process at zero temperature can be computed by \cite{Giorgini1998,Giorgini2000},
\begin{equation}
\gamma_B\left({\bf q}\right)=\frac{\pi}{2}\sum_{\textbf{p},\textbf{p}^\prime}|B_{\textbf{p}\textbf{p}^\prime}|^2
\delta(\epsilon_\textbf{q}-\epsilon_\textbf{p}-\epsilon_{\textbf{p}^\prime})\delta_{\textbf{q},\textbf{p}+\textbf{p}^\prime}. \label{Beliaev}
\end{equation}
Here, $B_{\textbf{p}\text{p}^\prime}$ is the matrix element associated with the scattering process wherein a quasiparticle having momentum ${\bf q}$ collides with the condensate creating two quasiparticles with momenta $\textbf{p}$ and $\textbf{p}^\prime$ [see Fig.~\ref{figure1}(a)], and the summation is performed over all possible states allowed by the energy and momentum conservation conditions specified in the $\delta$ function and $\delta_{\textbf{q},\textbf{p}+\textbf{p}^\prime}$ respectively.

To gain intuitions into how SOC affects Beliaev damping of low-energy excitations of BEC, we recall the classic result of the decay rate in a uniform one-component BEC with condensate density $n_0$, i.e., \cite{Liu1997,Giorgini1998,Giorgini2000}
\begin{equation}
\gamma_0=\frac{3q^5}{640\pi\hbar^3 m n_0}, \label{free}
\end{equation}
which exhibits the well-known $q^5$ scaling. Notice that the formula does not contain the interaction strength between bosonic atoms, rather, the role of interaction comes in only implicitly via $n_0$ \cite{Footnote2}. Equation (\ref{free}) also holds for a spin-$\frac{1}{2}$ BEC without SOC in the unpolarized phase \cite{Bhattacherjee2008,Bhattacherjee2014}. There, the density excitation is decoupled from the spin-density excitation, hence both the scattering matrix element and the conservation condition in Eq.~(\ref{Beliaev}) bear the same form as the one-component case apart from a renormalized interaction constant which, according to Eq.~(\ref{free}), does not alter the formal result. 

By contrast, as we will elaborate below, adding SOC will bring two fundamental differences: (i) The SOC breaks Galilean invariance, resulting in hybridized excitations in density and spin channels, so that the wavefunctions of low-energy quasiparticles and thus the Beliaev scattering matrix are strongly modified; (ii) The SOC renders a spatially anisotropic distribution of scattering states allowed by energy and momentum conservation. 

{\it Model Hamiltonian---}We consider a 3D spatially uniform BEC with a spin-orbit coupling along the $x$ axis. The relevant grand-canonical Hamiltonian is \cite{Zhai2012,Zhai2015} 
\begin{eqnarray}
K&=&\int d^3{\bf r}\hat{\psi}^\dagger\left({\bf r},t\right)\left(H_0-\mu N\right)\hat{\psi}\left({\bf r},t\right)\nonumber\\
&+&\frac{1}{4}\int d^3{\bf r}\left(g+g_{12}\right)\hat{n}^2\left({\bf r},t\right)+\left(g-g_{12}\right)\hat{S}_z^2\left({\bf r},t\right).\label{Ham}
\end{eqnarray}
Here, $\hat{\psi}^\dagger({\bf r},t)=(\hat{\psi}^\dagger_1,\hat{\psi}^\dagger_2)^T$ and $\hat{\psi}({\bf r},t)=(\hat{\psi}_1,\hat{\psi}_2)$ are the creation and annihilation operators for the two component bosonic atoms; $\hat{n}\left({\bf r},t\right)=|\hat{\psi}_1|^2+|\hat{\psi}_2|^2$ and $\hat{S}_z\left({\bf r},t\right)=|\hat{\psi}_1|^2-|\hat{\psi}_2|^2$ denote the total and spin density operators, respectively. The $g$ and $g_{12}$ denote the intra- and inter-species coupling constant, respectively, with $g\neq g_{12}$ in view of relevant experiments \cite{Lin2011}. The single-particle Hamiltonian $H_0$ contains a Zeeman term and an equal contribution of Rashba and Dresselhaus spin-orbital coupling in $x$- direction \cite{Lin2011,Ho2011,Li2012,Li2013,WuLi2015,ZhuChen2016}, i.e.,  
\begin{equation}
H_0=\frac{1}{2m}\left[(p_x-\hbar k_0\sigma_z)^2+p^2_{\perp}\right]+\frac{\hbar\Omega}{2}\sigma_x,\label{Single}
\end{equation}
where $m$ is the bare mass of bosonic atoms, $\sigma_i$ are standard Pauli matrices and $k_0$ labels the strength of SOC. Hamiltonian of such form has been recently realized in atomic setup \cite{Lin2009,Lin2011,Wangpj2012,Cheuk2012,Zhang2012,Ji2014,Li2016,Li2017} employing two counter-propagating Raman lasers, where $\Omega$ is the Raman coupling constant and $k_0$ is the momentum transfer between the lasers. We will moreover denote $G_1=(g+g_{12})n_0/4$ and  $G_1=(g-g_{12})n_0/4$ with $n_0$ the condensate density. 

Before continuing, let us briefly describe the ground state properties of Hamiltonian (\ref{Ham}). For $\hbar\Omega< 2\hbar^2k_0^2/m$, the single-particle dispersion $H_0$ exhibits degenerate double minima at momenta $p_x=\pm \hbar k_1$ with $k_1=k_0\sqrt{1-m^2(\hbar\Omega)^2/4\hbar^4k_0^4}$. In this regime, the ground state can exhibit a stripe phase \cite{Li2017} or a plane-wave phase \cite{Lin2011}. For $\hbar\Omega> 2\hbar^2k_0^2/m$, the single-particle dispersion features a global minimum at $k_1=0$ and is anisotropic. In this case, the condensate is in the zero-momentum phase \cite{Li2012} described by the familiar order parameter $
(\phi_{1}^0,\phi_{2}^0)=\sqrt{n_0/2}(1,-1)$. It is noteworthy that the mean field interaction will modify above boundary condition between the zero-momentum and plane-wave phases \cite{Li2012}, which becomes instead $\hbar\Omega=2\hbar^2k_0^2/m-4G_2$.

{\it  Beliaev damping in presence of SOC-- } Our goal is to investigate the Beliaev damping of the model system. We will assume $\hbar\Omega> 2\hbar^2k_0^2/m-4G_2$ \cite{Li2012} when the BEC is in the zero-momentum ground state phase, which represents the simplest case capturing essential effect of SOC on the dissipation of quasiparticles as mentioned earlier. 

We will first discuss the energy conservation condition in Eq.~(\ref{Beliaev}), since here a mean-field dispersion relation for the density excitation is sufficient. Writing $\Phi({\bf r})\equiv\langle \hat{\psi}({\bf r})\rangle=\phi_0({\bf r})+\delta\phi({\bf r})$, and noting that the relevant process involves mainly phonons in the low-momentum regime, we can write $
\epsilon_\textbf{k}=c_{\theta_{\bf k}} k$ for a phonon carrying momentum ${\bf k}$, with $k=|{\bf k}|$. Here, $c_{\theta_{\bf k}}$ is the sound velocity, which for Hamiltonian (\ref{Ham}) is found as $c_{\theta_{\bf k}}=\sqrt{1/\kappa m^*}$ \cite{LiEPL} where $\kappa^{-1}=2G_1$ is the compressibility \cite{Martone2012} and $m^*$ is the effective mass \cite{Liang2008} given by
\begin{equation}
\frac{m}{m^*}=1-\frac{2\hbar^{2}k_{0}^{2}\cos^{2}\theta_{\bf k}}{m\left(4G_2+\hbar\Omega\right)}. \label{EM}
\end{equation}
Here $\theta_{\bf k}$ measures the angle between the direction of momentum ${\bf k}$ of a quasiparticle and $x$-axis (along which SOC is applied). A crucial feature of the effective mass (\ref{EM}) is its spatial anisotropy: $m^*=m$ when ${\bf k}$ is perpendicular to the SOC direction while $m^*>m$ otherwise, as experimentally demonstrated \cite{Chen2015}. Notice that $m^*$ exhibits dependence on the spin-dependent interaction $G_2$, which for $G_2=0$ reduces to the result in Ref.~\cite{Martone2012}. Thus the energy conservation condition becomes strongly anisotropic, which for phonons takes the form $c_{\theta_\textbf{q}} q=c_{\theta_\textbf{p}} p+c_{\theta_\textbf{q-p}} |\textbf{q-p}|$ \cite{Footnote1}. 

The anisotropic energy condition results in an anisotropic distribution of scattering states contributing to Beliaev decay. To visualize this, we numerically solve the condition $\epsilon_\textbf{q}=\epsilon_\textbf{p}+\epsilon_\textbf{q-p}$. We will hereafter denote $\theta_{\bf q}=\theta$, i.e., the angle between the initial momentum ${\bf q}$ and $x$ direction. Figure~\ref{figure1}(b) presents the results of energetically allowed scattered momentum $\textbf{p}$ manifold for various $\theta$ on the $p_x-p_z$ plane ($q_y=p_y=0$ is taken). Interestingly, we see that the counterclockwise rotation of the manifold is accompanied by an increase of the manifold size with $\theta$, indicating anisotropic distribution of contributing states, in contrast to the SOC-free counterpart where the contour size stays invariant \cite{Beliaev2001E}. 

Next, we discuss the scattering matrix in Eq.~(\ref{Beliaev}), which instead requires \textit{beyond-mean-field} treatment. We follow the approach in Ref.~\cite{Giorgini1998}, which, by decomposing the total field operator $\hat{\psi}=\Phi+\tilde{\psi}$ where $\tilde{\psi}$ annihilates \textit{noncondensate} atoms and is treated perturbatively, allows for the account of couplings between Bogoliubov quasiparticles and noncondensate atoms. In this framework \cite{Supp}, the matrix element $B_{{\bf p}{\bf p^{\prime}}}$ in terms of usual Bogoliubov amplitudes $u(v)$ reads $B_{\textbf{p}\textbf{p}'}=\tilde{B}_{\textbf{p}\textbf{p}'}+\tilde{B}_{\textbf{p}'\textbf{p}}$, with 
\begin{widetext}
\begin{eqnarray}
\tilde{B}_{\textbf{p}\textbf{p}'}&=\sqrt{\frac{n_{0}}{2V}}\sum_{\alpha=1,2}(-1)^{\alpha+1}\big\{[g(2u_{\alpha,\textbf{p}}
v_{\alpha,\textbf{p}^\prime}+u_{\alpha,\textbf{p}}u_{\alpha,\textbf{p}^\prime})
+g_{12}(u_{\bar{\alpha},\textbf{p}}v_{\bar{\alpha},\textbf{p}^\prime}-u_{\bar{\alpha},\textbf{p}}u_{\alpha,\textbf{p}^\prime}
-u_{\alpha,\textbf{p}}v_{\bar{\alpha},\textbf{p}^\prime})]u_{\alpha,\textbf{q}}\nonumber\\
&+[g(2u_{\alpha,\textbf{p}}v_{\alpha,\textbf{p}^\prime}
+v_{\alpha,\textbf{p}}v_{\alpha,\textbf{p}^\prime})+g_{12}(u_{\bar{\alpha},\textbf{p}}v_{\bar{\alpha},\textbf{p}^\prime}
-v_{\bar{\alpha},\textbf{p}}v_{\alpha,\textbf{p}^\prime}
-v_{\alpha,\textbf{p}}u_{\bar{\alpha},\textbf{p}^\prime})]v_{\alpha,\textbf{q}}\big\}.\label{Prefactor}
\end{eqnarray}
\end{widetext}
We will now take an experimental viewpoint by describing Eq.~(\ref{Prefactor}) in terms of the dynamic structure factors \cite{PineBook}, as inspired by Ref.~\cite{Beliaev2002E}. In cold atom experiments, the dynamic structure factor can be directly measured by means of Bragg spectroscopy \cite{Kurn1999,Steinhauer2002,Du2010,Ozeri2005} or {\it in situ} imaging \cite{Hung2011,Hung2013}, as in recent studies of SOC BECs \cite{Khamehchi2014,Chen2015,Ha2015}, which gives experimental access to the Bogliubov amplitudes $u(v)$ \cite{Brunello2000,Vogels2002}. A SOC BEC has two types of dynamic structure factor \cite{Li2013,Martone2012}, i.e., the density- and spin-density dynamic structure factors, describing the system response to the density- and spin-density perturbations, respectively. Formally, the density dynamic structure factor is given by  
$S_d\left({\bf q},\omega \right)=N^{-1}\sum_n |\langle 0 | \rho_{{\bf q}} |n\rangle|^2\delta(\omega-\omega_{n0})$ where $\rho_{{\bf q}} =\sum_ie^{i {\bf q}\cdot {\bf x}_i}$ is the density
operator with momentum ${\bf q}$ and $\omega_{n0}$ is the 
excitation frequency of the $n$-th state, while the spin density dynamic structure factor is $S_s\left({\bf q},\omega \right)=N^{-1}\sum_n |\langle 0 | s_{{\bf q}} |n\rangle|^2\delta(\omega-\omega_{n0})$ with $s_{{\bf q}} =\sum_i\sigma_{zi}e^{i {\bf q}\cdot {\bf x}_i}$ being the standard spin density
operator. The static density and spin density structure factor are thus $S_{d(s)}\left({\bf q}  \right)=\int d\omega S_{d(s)}\left({\bf q},\omega \right)$, with $S_d+S_s=1$. 

Without SOC, the density and spin-density excitations of a spin-$\frac{1}{2}$ BEC are decoupled, so that an external density perturbation $\delta\hat
\rho$ acting on BECs only induces a density response in form of the density dynamic structor factor. Instead, due to the absence of Galilean invariance in a SOC BEC, a density perturbation along the $x$-direction in the system, which formally corresponds to a gauge transformation $e^{i  q_x x}$ \cite{PineBook}, will concomitantly induce a velocity dependent Zeeman-energy term $-q_x\hbar k_0\sigma_z$, resulting in generations of both density- and spin-density responses. Thus the mechanism of SOC affecting the scattering matrix and thereby Beliaev damping by coupling excitations in the density- and spin- channels can be demonstrated through the effect of SOC on the static structure factors.

Figures \ref{figure1}(c) and (d) compare $S_d\left({\bf q}\right)$ [Fig.~\ref{figure1}(c)]
and $S_s\left({\bf q}\right)$ [Fig.~\ref{figure1}(d)] for various SOC strength $k_0$, taking $\theta=0$. Without SOC, it is well known that $S_d$ asymptotically approaches unity at large momenta while $S_s$ is pined to zero (see blue solid curves). By contrast, the most prominent feature in presence of SOC is that $S_s$ becomes finite at all momenta, signaling the coupling of density and spin-density excitations. In particular, at large momenta, both $S_d$ and $S_s$ become to unanimously approach $1/2$ [see insets of Figs.~\ref{figure1}(c) and (d)]. Such different asymptotic behavior compared to the SOC-free case can be analytically understood as follows: The static structure factor can be written as $S_d=N^{-1}\left|\sum_{\alpha=1,2}\sqrt{n}_{i0}\left(u_{\alpha\bf q}+v_{\alpha\bf q}\right)\right|^2$ and $S_s=N^{-1}\left|\sum_{\alpha=1,2}\sqrt{n}_{\alpha0}{\text sgn}(\alpha)\left(u_{\alpha\bf q}+v_{\alpha\bf q}\right)\right|^2$ with ${\it sgn}(1)=-{\it sgn}(2)=1$. For $q\rightarrow 
\infty$, when $k_0=0$ we have $u_{1{\bf {\bf q}}}=v_{1{\bf {\bf q}}}=1/\sqrt{2}$ and $u_{2{\bf {\bf q}}}=v_{2{\bf {\bf q}}}=0$, instead, when $k_0\neq 0$ we have $u_{1{\bf {\bf q}}}\rightarrow1$, $v_{1{\bf {\bf q}}}\rightarrow0$, $u_{2{\bf {\bf q}}}=v_{2{\bf {\bf q}}}=0$. At small momenta, we see that the increase rate $S_s$ enhances with $k_0$ as expected. For arbitrary momenta, the Bogoliubov amplitudes $u$ and $v$ in Eq.~(\ref{Prefactor}) can be related to $S_d$ and $S_s$ as \cite{Coefficient}
\begin{eqnarray}
u_{\alpha,\textbf{q}}&=&\frac{f+(-1)^{\alpha+1}2\beta_{q}[(\sqrt{S_{d}}+(-1)^{\alpha+1}\sqrt{S_{s}})^{2}+1]}{4\sqrt{2}(\sqrt{S_{d}}+(-1)^{\alpha+1}\sqrt{S_{s}})\beta_{q}},\nonumber\\
v_{\alpha,\textbf{q}}&=&\frac{-f+(-1)^{\alpha+1}2\beta_{q}[(\sqrt{S_{d}}+(-1)^{\alpha+1}\sqrt{S_{s}})^{2}-1]}{4\sqrt{2}(\sqrt{S_{d}}+(-1)^{\alpha+1}\sqrt{S_{s}})\beta_{q}}. \nonumber
\end{eqnarray}
Thus by measuring the dynamic structure factor and hence accessing Bogoliubov amplitudes \cite{Brunello2000,Vogels2002}, one can access the matrix element in Eq.~(\ref{Prefactor}) for the SOC BEC, along the lines of the Beliaev damping experiments in the one-component BEC \cite{Beliaev2002E}.

Finally, in performing the summation in Eq.~(\ref{Beliaev}), we will assume all momenta are along the same direction \cite{Beliaev2001E,Ozeri2005}, i.e., $\theta_{\bf q}=\theta_{\bf p}=\theta_{\bf {p}-\bf {q}}=\theta$, as collisions at zero temperature dominantly occurs in the low-momentum regime where the energy and momentum conservation conditions require the scattered momentum ${\bf p}$ be parallel with the initial momentum ${\bf q}$. This way, straightforward evaluation gives \cite{Supp} 
\begin{equation}
\!\!\!\!\!\gamma_B=\gamma_0\left[1-\frac{2\hbar^3\Omega k_0^2\cos^2\theta}{m (4G_2+\hbar\Omega)^2}\right]^2 \sqrt{1+\chi_M\frac{\hbar^2k_0^2}{m} \sin^2\theta}. \label{BF}
\end{equation}
Here $\chi_M$ is the spin polarizability susceptibility \cite{LiEPL,Martone2012} which takes the form
\begin{equation}
\chi_M=\frac{2}{\left(\hbar\Omega+4G_{2}\right)-2\hbar^{2}k_{0}^{2}/m}.\label{SPS}
\end{equation}

Equation (\ref{BF}) is the key result of this Letter. Apparently, $\gamma_B$ for $k_0=0$ reduces to $\gamma_0$ of the SOC-free counterpart [see Eq.~(\ref{free})]. While maintaining the familiar $q^5$ dependence [see Fig.~\ref{figure2}(a)], $\gamma_B$ displays following distinguishing features in contrast to $\gamma_0$: 

\begin{figure}[!ht]
\centering
\includegraphics[width=1\columnwidth]{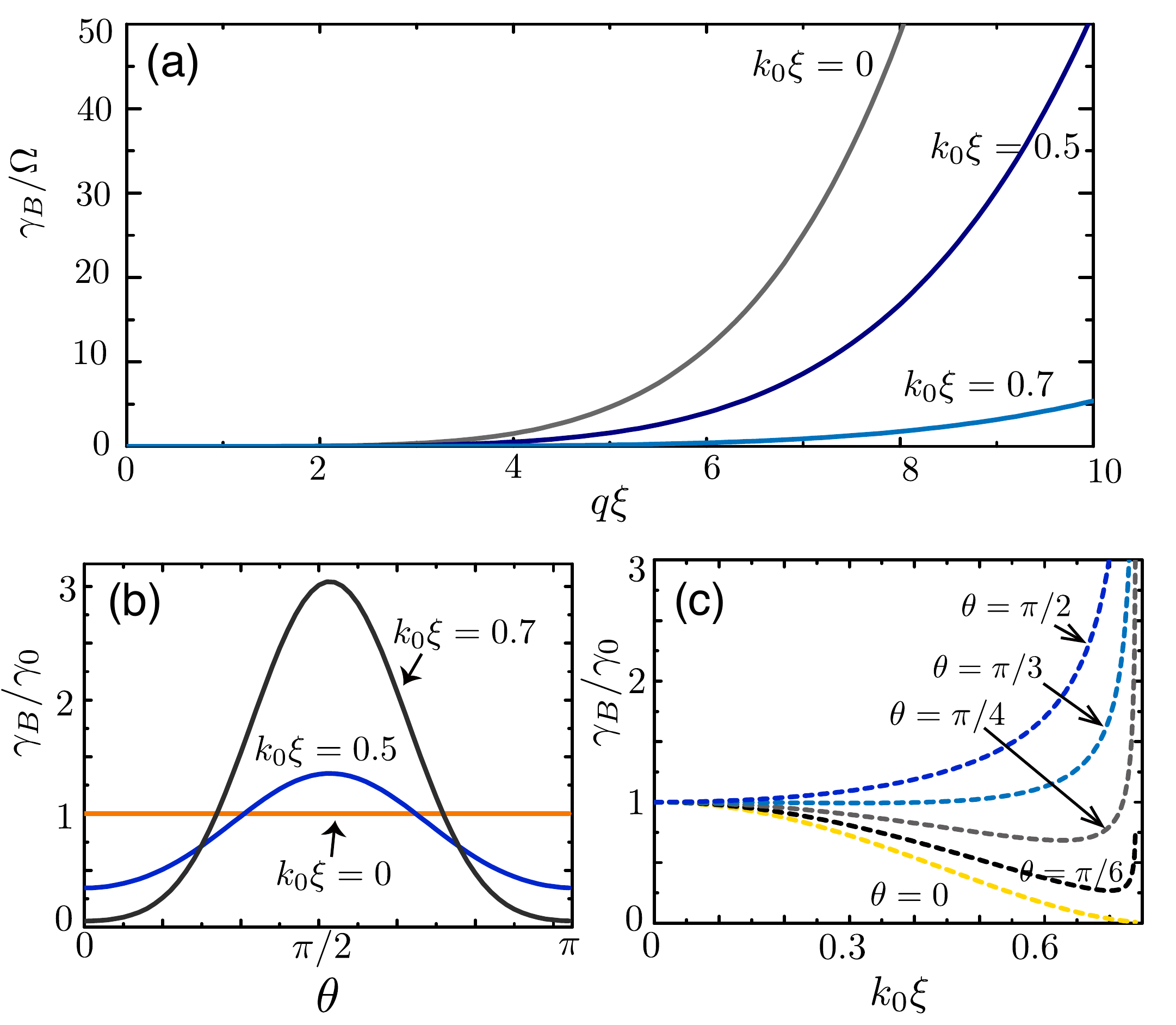}
\caption{ (Color online). The Beliaev damping rate $\gamma_B$ as a function of (a) the modulus of momentum $q=|{\bf q}|$, fixing $\theta=0$; (b) the angle $\theta$, taking $q\xi=0.8$; (c) the SOC strength $k_0$. In all plots, we take $G_1/\hbar\Omega = 0.1$; $G_2/\hbar\Omega = 0.025$.   }\label{figure2}
\end{figure}

(i) $\gamma_B$ is \textit{explicitly interaction-dependent}, which comes in only via $g-g_{12}$ (contained in $G_2$) characterizing the strength of spin-density interaction [see Eq.~(\ref{Ham})]. This presents a clear manifestation of the coupled density and spin-density excitations due to SOC on phonon dissipations. Interestingly, $\gamma_B$ at $\theta \neq 0$ exhibits a characteristic divergence at the critical point $4G_2+\hbar\Omega=2\hbar^2 k_0^2/m$, which is just the aforementioned phase boundary between the zero-momentum and plane wave phases. This divergence of $\gamma_B$ comes from the divergence of spin polarizability susceptibility $\chi_M$. As discussed earlier, a density perturbation due to presence of SOC is necessarily accompanied by a perturbation $\sim \sigma_z$. This induces a system response in form of the spin polarizability susceptibility, which has been shown to be able to distinguish the unpolarized zero-momentum phase and the spin-polarized plane-wave phase. Note that the measurement of spin polarizability for the considered SOC BEC has been recently reported \cite{Zhang2012}. By contrast, $\gamma_B$ at $\theta = 0$ always stays finite. This can be understood by noticing the effective mass along the SOC direction diverges [see Eq.~(\ref{EM})] giving rise to the so called phonon softening \cite{Chen2015}, which at the phase boundary effectively cancels the divergence of $\chi_M$. We note that the determination of condensate density $n_0$ in Eq.~(\ref{BF}) relies on the density interaction constant $g+g_{12}$ \cite{Zheng2013}.

 (ii) $\gamma_B$ is strongly anisotropic depending on the angle between initial momentum ${\bf q}$ of quasiparticle and SOC direction, which can be understood in terms of the SOC-induced anisotropic effective mass. In fact, Eq.~(\ref{BF}) can be cast into a more transparent form by ignoring $G_2$, i.e., 
 \begin{equation}
\gamma_B=\Big(\frac{3q^5}{640\pi\hbar^3 m^* n_0}\Big)\frac{m}{m^*} \sqrt{1+\chi_M\frac{\hbar^2k_0^2}{m} \sin^2\theta}.\label{BF1}
\end{equation}
Thus, for a fixed SOC strength $k_0$, the decay of quasiparticle is most significant when ${\bf q}$ is perpendicular to the SOC direction, but is strongly suppressed when the two are parallel [see Fig.~\ref{figure2}(b)]. In addition, when increasing SOC [see Fig.~\ref{figure2}(c)], the decay along the SOC direction is increasingly suppressed while that in the perpendicular direction is enhanced, although for other directions, $\gamma_B$ is generally nonmonotonic with respect to $k_0$.
 
 {\it Concluding discussions---} Summarizing, we have shown how the effect of SOC can manifest itself in the Beliaev damping of low-energy excitations of a BEC, even when the ground state is in the zero momentum phase, and the essential features such as anisotropy and the dependence on the spin-density should also be seen in the plane-wave phase and the stripe phase. In the latter phases, since the ground-state wavefunctions and the single-particle dispersions already bear clear signatures of SOC effect (unlike the zero-momentum phase), the explorations of the unique features of quasiparticle decay there remain an open challenge. In addition, our analysis connects the damping rate with the presently detectable dynamical structure factors, thus opens possibility for experimental access, e.g., by means of Bragg spectroscopy. While many-body quantum systems with SOC have been intensively studied within the mean-field framework, observing the Beliaev damping in a SOC BEC would present an important step toward revealing the interplay between the non-Abelian gauge fields and the beyond mean effects. 
\bigskip

{\it Acknowledgement---}We would like to thank Hua Chen, Chao Gao, Xianlong Gao, Ying Hu, and Wei Yi for stimulating
discussions. This work is financially supported by the National Natural Science Foundation of China (Grants Nos. 11274315 and 11774316) and Zhejiang Provincial Natural Science Foundational of China under Grant No. LQ13A040005.

\bibliography{myr}

\begin{thebibliography}{65}
\bibitem [{Foo({\natexlab{a}})}]{Footnote2}%
  \BibitemOpen
  \href@noop {} {} ({\natexlab{a}}),\ \bibinfo {note} {for uniform
  one-component BEC, the $n_0$ can be written as $n_0=n-n_{ex}$ with $n$ and
  $n_{ex}/n=(8/3\sqrt{\pi})(na_s^3)^{1/2}$ being the total density and quantum
  depletion respectively. The Beliaev damping rate in Eq. (2) is implicitly
  dependent on the interaction constant $g=4\pi\hbar^2a_s/m$ through
  $n_0=n-n_{ex}$.}\BibitemShut {Stop}%
\bibitem [{Foo({\natexlab{b}})}]{Footnote1}%
  \BibitemOpen
  \href@noop {} {} ({\natexlab{b}}),\ \bibinfo {note} {for the analysis of the
  scattering matrix, the phonon dispersion needs to be modified by including a
  nonlinear term \cite{Giorgini1998}, i.e., $\epsilon_\textbf{q}=c_{\theta_{\bf
  q}} q+d_{\theta_{\bf q}} q^3$.}\BibitemShut {Stop}%
\end{thebibliography}
\end{document}